%% file: ms.tex
\documentclass[conference, doublecolumn, hidelinks, xcolor=dvipsnames]{IEEEtran}

\pdfoutput=1
\IEEEoverridecommandlockouts
\usepackage{microtype}
\usepackage{acronym}
\usepackage{amsmath}
\usepackage{amssymb}
\usepackage[english]{babel}
\usepackage{cite}
\usepackage[shortlabels,inline]{enumitem}
\usepackage{graphicx}
\usepackage{hyperref}
\usepackage{cleveref}
\usepackage[utf8]{inputenc}
\usepackage{lipsum}
\usepackage{siunitx}
\usepackage[dvipsnames]{xcolor}
\usepackage[T1]{fontenc}
\usepackage{threeparttable}

\newtoggle{Arxiv}

\toggletrue{Arxiv}

\newcommand{\version}[2]{%
\iftoggle{Arxiv}{%
#1
}
{#2}
}
\allowdisplaybreaks

\newcommand{\Figure}[1]{Fig.~\ref{#1}}

\newcommand{\Equation}[1]{\eqref{#1}}

\newcommand{\Table}[1]{Table~\ref{#1}}

\newcommand{\Section}[1]{Section~\ref{#1}}

\newcommand{\Sections}[2]{Sections~\ref{#1}~and~\ref{#2}}

\input{mathCommandsAndOperators.tex}
\input{acronyms.tex}

\makeatletter
\long\def\@makecaption#1#2{\ifx\@captype\@IEEEtablestring%
    \footnotesize\begin{center}{\normalfont\footnotesize #1}\\
        {\normalfont\footnotesize\scshape #2}\end{center}%
    \@IEEEtablecaptionsepspace
    \else
    \@IEEEfigurecaptionsepspace
    \setbox\@tempboxa\hbox{\normalfont\footnotesize {#1.}~~ #2}%
    \ifdim \wd\@tempboxa >\hsize%
    \setbox\@tempboxa\hbox{\normalfont\footnotesize {#1.}~~ }%
    \parbox[t]{\hsize}{\normalfont\footnotesize \noindent\unhbox\@tempboxa#2}%
    \else
    \hbox to\hsize{\normalfont\footnotesize\hfil\box\@tempboxa\hfil}\fi\fi}
\makeatother

\newcommand{\scaleSection}{\vspace{-0.205cm}}
\newcommand{\scaleSubsection}{\vspace{-0.155cm}}
\newcommand{\scaleSubsubsection}{\vspace{-0cm}}
\newcommand{\scaleSectionBelow}{\vspace{-0.175cm}}
\newcommand{\scaleSubsectionBelow}{\vspace{-0.095cm}}
\newcommand{\scaleSubsubsectionBelow}{\vspace{-0.025cm}}
\newcommand{\scaleAlign}{\vspace{-0.185cm}}
\renewcommand{\baselinestretch}{0.93}
\begin{document}
\bstctlcite{IEEEexample:BSTcontrol}
\vspace{-0.4cm}
\title{Closed Loop Molecular Communication Testbed:\vspace{-0.05cm} Setup, Interference Analysis, and Experimental \vspace{-0.05cm}Results\vspace{-0.45cm}}
\author{
\IEEEauthorblockN{Lukas Brand$^\star$, Maike Scherer$^\star$, Teena tom Dieck, Sebastian Lotter, \\Maximilian Schäfer, Andreas Burkovski, Heinrich Sticht, Kathrin Castiglione, and Robert Schober
\thanks{$^\star$ Co-first authors.}}\IEEEauthorblockA{\small Friedrich-Alexander-Universit\"at Erlangen-N\"urnberg, Erlangen, Germany}\vspace{-0.98cm}}
\maketitle

\thispagestyle{plain}
\pagestyle{plain}
\begin{abstract}
\input{abstract}
\end{abstract}
\setlength{\belowdisplayskip}{2pt}
\setlength{\belowdisplayshortskip}{2pt}
\acresetall
\scaleSection
\section{Introduction }\label{intro}
\scaleSectionBelow
\input{introduction}
\scaleSection
\section{Signaling Molecule: GFPD and its Properties}\label{sec:GFPD}
\scaleSectionBelow
\vspace{0.1cm}
\input{GFPD}
\scaleSection
\section{Testbed Setup}\label{testbed}
\scaleSectionBelow
\input{testbed}
\vspace{-0.05cm}
\scaleSection
\section{Communication Link Characterization}\label{characterization}
\scaleSectionBelow
\input{characterization}
\scaleSection
\section{Detection Methods}\label{Estimation_and_Detection}
\scaleSectionBelow
\input{performance}
\scaleSection
\section{End-To-End Communication Performance}\label{evaluation}
\scaleSectionBelow
\input{evaluation}
\scaleSection
\section{Conclusion}\label{sec:conclusion}
\scaleSectionBelow
\input{conclusion}
\vspace{-0.1cm}
\scaleSection
\renewcommand{\baselinestretch}{0.93}
\bibliographystyle{IEEEtran}
\bibliography{literature}
\scaleSectionBelow
\renewcommand{\baselinestretch}{0.93}
\scaleSubsection
\vspace{0.1cm}
\section*{Acknowledgment}
\scaleSectionBelow
{\small
We thank Prof. Stefan Jakobs (Max Planck Institute for Biophysical Chemistry, Göttingen, Germany) for providing a plasmid encoding Dreiklang.
}
\end{document}

%% file: mathCommandsAndOperators.tex
\newcommand{\con}[0]{C_{\mathrm{GFPD}}} 
\newcommand{\volTube}[0]{V_{\mathrm{T}}} 

\newcommand{\Tg}[0]{T_{\mathrm{G}}} 
\newcommand{\Ts}[0]{T_{\mathrm{S}}} 
\newcommand{\Ti}[0]{T_{\mathrm{I}}} 
\newcommand{\bitest}[2][]{\hat{b}_{#1}[#2]} 
\newcommand{\bit}[1]{b[#1]} 
\newcommand{\rec}[1]{d[#1]} 
\newcommand{\recDiff}[1]{d'[#1]} 

\newcommand{\threshR}[0]{\xi_{\mathrm{R}}} 
\newcommand{\threshD}[0]{\xi_{\mathrm{D}}} 
\newcommand{\nPilots}[0]{L_{\mathrm{P}}} 
\newcommand{\nSymb}[0]{N_{\mathrm{Sym}}} 
\newcommand{\idxSym}[1]{\mathcal{K}^{\mathrm{P}}_{#1}} 


%% file: acronyms.tex
\acrodef{1D}[1-D]{one-dimensional}
\acrodef{2D}[2-D]{two-dimensional}
\acrodef{3D}[3-D]{three-dimensional}
\acrodef{MC}{molecular communication}
\acrodef{OOK}{ON-OFF keying}
\acrodef{ISI}{inter-symbol interference}
\acrodef{IUI}{inter-user interference}
\acrodef{IR}{impulse response}
\acrodef{ML}{maximum likelihood}
\acrodef{wlog}[w.l.o.g.]{without loss of generality}
\acrodef{BER}{bit error rate}
\acrodef{SER}{symbol error rate}
\acrodef{BSC}{Binary Symmetric Channel}
\acrodef{CDF}{cumulative density function}
\acrodef{UCA}{uniform concentration assumption}
\acrodef{AWGN}{Additive White Gaussian Noise}
\acrodef{PBS}{particle-based simulation}
\acrodef{MLSE}{maximum likelihood sequence estimator}
\acrodef{VE}{viterbi equalizer}
\acrodef{MLE}{maximum likelihood estimator}
\acrodef{CIR}{channel impulse response}
\acrodef{CIRs}{channel impulse responses}
\acrodef{SNR}{signal-to-noise ratio}
\acrodef{SDMA}{Space Division Multiple Access}
\acrodef{MCDMA}{Molecular Code Division Multiple Access}
\acrodef{ADMA}{Amplitude-Division Multiple Access}
\acrodef{MTDMA}{Molecular Time Division Multiple Access}
\acrodef{MDMA}{Molecular Division Multiple Access}
\acrodef{MIMO}{multiple-input multiple-output}
\acrodef{TDMA}{Time Division Multiple Access}
\acrodef{CDMA}{Code Division Multiple Access}
\acrodef{CSI}{channel state information}
\acrodef{TX}{transmitter}
\acrodef{TXs}{transmitters}
\acrodef{RX}{receiver}
\acrodef{RXs}{receivers}
\acrodef{ARE}{area rate efficiency}
\acrodef{w.r.t.}{with respect to}
\acrodef{GFP}{green fluorescent protein}
\acrodef{GFPD}{\textit{green fluorescent protein variant "Dreiklang"}}
\acrodef{EX}{eraser}
\acrodef{SM}{signaling molecule}
\acrodef{SMs}{signaling molecules}
\acrodef{UV}{ultraviolet}
\acrodef{LB}{lysogeny broth}
\acrodef{OD600}{optical density at 600 nm}
\acrodef{lac operon}[]{lactose operon}
\acrodef{IPTG}{Isopropyl-$\beta$-D-thiogalactopyranosid}
\acrodef{IMAC}{immobilised metal chelate affinity chromatography}
\acrodef{CV}{column volume}
\acrodef{BCA}{bicinchoninic assay}
\acrodef{SDS-PAGE}{sodium dodecyl sulfate polyacrylamide gel electrophoresis}
\acrodef{YFP}{yellow fluorescent protein}
\acrodef{FPs}{Fluorescent Proteins}
\acrodef{RSFPs}{Reversable switchable fluorescent proteins}
\acrodef{LED}{light emitting diode}
\acrodef{PWM}{pulse width modulation}
\acrodef{GFP}{green fluorescent protein}
\acrodef{FP}{fluorescent protein}
\acrodef{IoBNT}{Internet of Bio-Nano-Things}
\acrodef{FCQD}{fluorescent carbon quantum dot}
\acrodef{CSK}{concentration shift keying}
\acrodef{SPION}{superparamagnetic iron oxide nanoparticle}
\acrodef{BD}{basic threshold detection}
\acrodef{DD}{differential threshold detection}
\acrodef{UV}{ultraviolet}
\acrodef{NMSED}{normalized minimum squared Euclidean distance}

%% file: abstract.tex
In this paper, we present a fluid-based experimental molecular communication (MC) testbed that, similar to the human cardiovascular system, operates in a closed circuit tube system. The proposed system is designed to be biocompatible, resource-efficient, and controllable from outside the tube. As signaling molecule, the testbed employs the \textit{green fluorescent protein variant "Dreiklang"} (GFPD). GFPDs can be reversibly switched via light of different wavelengths between a bright fluorescent state and a less fluorescent state. Hence, this property allows for writing and erasing information encoded in the state of the GFPDs already present in the fluid via radiation from outside the tube. The concept of modulating the GFPDs existing in the channel at the transmitter for information transmission, instead of releasing new molecules, is a form of \textit{media modulation}.
In our testbed, due to the closed loop setup and the long experiment durations of up to 250 min, we observe new forms of inter-symbol interferences (ISI), which do not occur in short experiments and open loop systems. In particular, up to four different forms of ISI, namely channel ISI, inter-loop ISI, offset ISI, and permanent ISI, occur in the considered system.
To mitigate inter-loop ISI and offset ISI, we propose a light based eraser unit. We experimentally demonstrate reliable information transmission in our testbed achieving error-free transmission of 500 bit at a data rate of 6 $\textrm{bit}\, \textrm{min}^{\boldsymbol{-1}}$ based on a sub-optimal low-complexity detection scheme.

%% file: introduction.tex
Synthetic \ac{MC} refers to information transmission based on signaling molecules \cite{nakano2013molecular}. This paradigm makes it possible to design systems capable of operating in environments unsuitable for electromagnetic wave-based communication, such as the human cardiovascular system. For the latter, the \ac{IoBNT} has been proposed, a communication network potentially integrating many in-body nanodevices that utilize \ac{MC} to exchange information \cite{akyildiz2015internet}.

So far, such systems exist only in theory, i.e., eHealth products and applications based on \ac{MC} are not available, yet.
To drive research towards first applications, several \ac{MC} testbeds have been developed in recent years (see \cite{lotter2023experimental, Lotter2023testbedII} for a comprehensive overview of existing testbeds). For example, the authors in \cite{lee2023internet} propose a $2 \times 2$ fluid-based \ac{MIMO} system as a prototype for the \ac{IoBNT}.

However, the prototype in \cite{lee2023internet} and all other fluid-based \ac{MC} testbeds operating in tubes, except that in \cite{tuccitto2017fluorescent}, are open loop. Open loop refers to systems where the background fluid used in the experiment can enter or leave the system during operation. As a result, in these testbeds, the signaling molecules are added to the tube system at one point, e.g., via an injection, used \textit{once} to transmit information, and then are collected as waste together with the flowing out background fluid at the end of the tube. There are two major drawbacks to such systems: First, long-term transmission experiments generate a lot of waste. Second, many of the intended applications occur in a closed loop environment, e.g., the human cardiovascular system, and cannot be accurately simulated in open loop testbeds.

Therefore, experimental \ac{MC} system are required that can operate in closed loop. Currently, there is only one \ac{MC} testbed that features a closed loop \cite{tuccitto2017fluorescent}. In \cite{tuccitto2017fluorescent}, a fluid is pumped in a loop system while fluorescent particles are injected, detected, and then diluted. Hence, the applicability of the testbed in \cite{tuccitto2017fluorescent} for long experiment durations is limited because the signaling molecules can be used only once, which is not resource efficient.

In this context, we propose the first experimental closed loop \ac{MC} system in which reversibly switchable signaling molecules are reused multiple times, enabled by molecular \textit{media modulation} \cite{Brand2022MediaModulation, brand2023switchable}.
In media modulation, the state of signaling molecules already present in the \textit{medium} is changed to transmit information, i.e., no additional molecules are injected or removed from the system during operation.
As signaling molecule, we adopt the biocompatible \ac{GFPD} \cite{richards2003safety, brakemann2011reversibly}. \ac{GFPD} molecules can be reversibly switched between a bright fluorescent ON state and a less fluorescent OFF state via light stimuli of different wavelengths \cite{brakemann2011reversibly}. This allows for writing and erasing information employing an optical \ac{TX} and \ac{EX} in our testbed, respectively. The state of the \acp{GFPD}, i.e., the transmitted information, is read out via fluorescence detection at the \ac{RX}.

The main contributions of this paper are as follows:\vspace{-0.05cm}
\begin{itemize}
\item We utilize light-based media modulation to transmit information in a closed loop tube system using the states of \ac{GFPD} molecules. This form of media modulation enables to operate \ac{TX}, \ac{EX}, and \ac{RX} outside of the tube. As a result, they do not interfere with the propagation of the signaling molecules inside the tube; a feature that is beneficial, e.g., for healthcare applications. 
\item
Because of the closed loop and long experiment durations, the system is affected by new forms of \ac{ISI}.
These forms of \ac{ISI} become apparent on different time scales and are characterized in this work.
Furthermore, we discuss and experimentally implement suitable \ac{ISI} mitigation methods. Moreover, we implement two different detection schemes and test their robustness to \ac{ISI}.
\item We experimentally prove that error-free information transmission of $500 \,\si{bit}$ at $6 \,\si{ \bit \per\minute}$ and at $2 \,\si{ \bit \per\minute}$ is feasible in a closed loop tube system. The corresponding experiments take $83 \,\si{\minute}$ and $250 \,\si{\minute}$, respectively, during which $9 \, \si{\milli\liter}$ of \ac{GFPD} solution is continuously reused.
\end{itemize}

A preliminary and simplified version of the testbed was briefly presented as part of \cite{brand2023switchable}. In \cite{brand2023switchable}, the experiment duration was only $3 \,\si{\minute}$, which did not allow to i) evaluate the resource efficiency of the testbed, ii) characterize the \ac{ISI}, and iii) experimentally determine the \ac{SER} for different detection schemes, which are new contributions of this work.

The remainder of this paper is organized as follows. In \Sections{sec:GFPD}{testbed}, we describe \ac{GFPD} and the experimental \ac{MC} setup, respectively. In \Section{characterization}, we characterize the testbed and the different forms of \ac{ISI}. The employed detection methods are introduced in \Section{Estimation_and_Detection}, and in \Section{evaluation}, we evaluate the end-to-end communication performance. Finally, \Section{sec:conclusion} concludes the paper and outlines topics for future~work.

%% file: GFPD.tex
\ac{GFPD} is a \textit{photoswitchable} and \textit{biocompatible} \ac{GFP} introduced in \cite{brakemann2011reversibly}. In the following, we discuss these two properties which are key for the proposed \ac{MC} system.

\scaleSubsection
\subsection{Photoswitchability}\label{sec:GFPD_photoswitchable}
\scaleSubsectionBelow
In general, \textit{photoswitchability} means that the properties of a molecule can be changed upon exposure to light of specific wavelengths. In the case of \ac{GFPD}, photoswitchability allows for a reversible switching of its fluorescence between two\footnote{In fact, a third state exists, which we denote as \textit{equilibrium state}. In the absence of light, \ac{GFPD} may switch to this state, which shows a similar fluorescence as the ON state. As the equilibrium state's fluorescence is only slightly higher, in this work, for simplicity of presentation, we do not differentiate between the ON state and the equilibrium state.} distinct states, namely the ON and OFF states, which can be controlled by irradiation of light with different wavelengths.
In particular, \ac{GFPD} may be switched to the ON and OFF states upon irradiation with light of wavelengths $\lambda_\mathrm{ON} = 365 \, \si{\nm}$ and $\lambda_\mathrm{OFF} = 405 \, \si{\nm}$, respectively. The success of the switching depends on the intensity of the irradiation. Furthermore, \ac{GFPD} can also spontaneously switch back to the ON state due to thermal relaxation. The thermal relaxation of \ac{GFPD} to the ON state has a half-life of $T_{1/2} = 600 \,\si{\second}$ at room temperature \cite{brakemann2011reversibly}.
Only \acp{GFPD} in the ON state show a high fluorescence, i.e., emit light at a wavelength of $\lambda_\mathrm{E} = 529 \, \si{\nm}$ after being excited by light of wavelength $\lambda_\mathrm{T} = 500 \, \si{\nm}$.
After multiple photoswitching cycles, photobleaching of \ac{GFPD} may occur \cite{brakemann2011reversibly}. Here, photobleaching refers to the chemical modification of the \ac{GFPD} upon exposure to light of high energy. The effect of photobleaching is a gradual diminishing of the fluorescence intensity upon repeated irradiation which can be interpreted as \ac{GFPD} degradation.
The authors in \cite{brakemann2011reversibly} show that, for \ac{GFPD}, the switching efficiency, robustness towards photobleaching, and fluorescence intensity are high, which makes \ac{GFPD} well suited for the use in our testbed.

\scaleSubsection
\subsection{Biocompatibility}\label{sec:GFPD_bio}
\scaleSubsectionBelow
Like all \acp{GFP}, \ac{GFPD} is \textit{biocompatible} \cite{richards2003safety}. This property allows the use of \ac{GFP}-based proteins in a wide range of living organisms without causing significant harm or disruption to their normal functions \cite{richards2003safety}. The biocompatibility makes \acp{GFPD} suitable for \textit{in vivo} usage.

%% file: testbed.tex
\begin{figure}[!tbp]
\centering
  \includegraphics[width=0.9\columnwidth, trim = 0.5cm 2.1cm 13cm 0cm, clip]{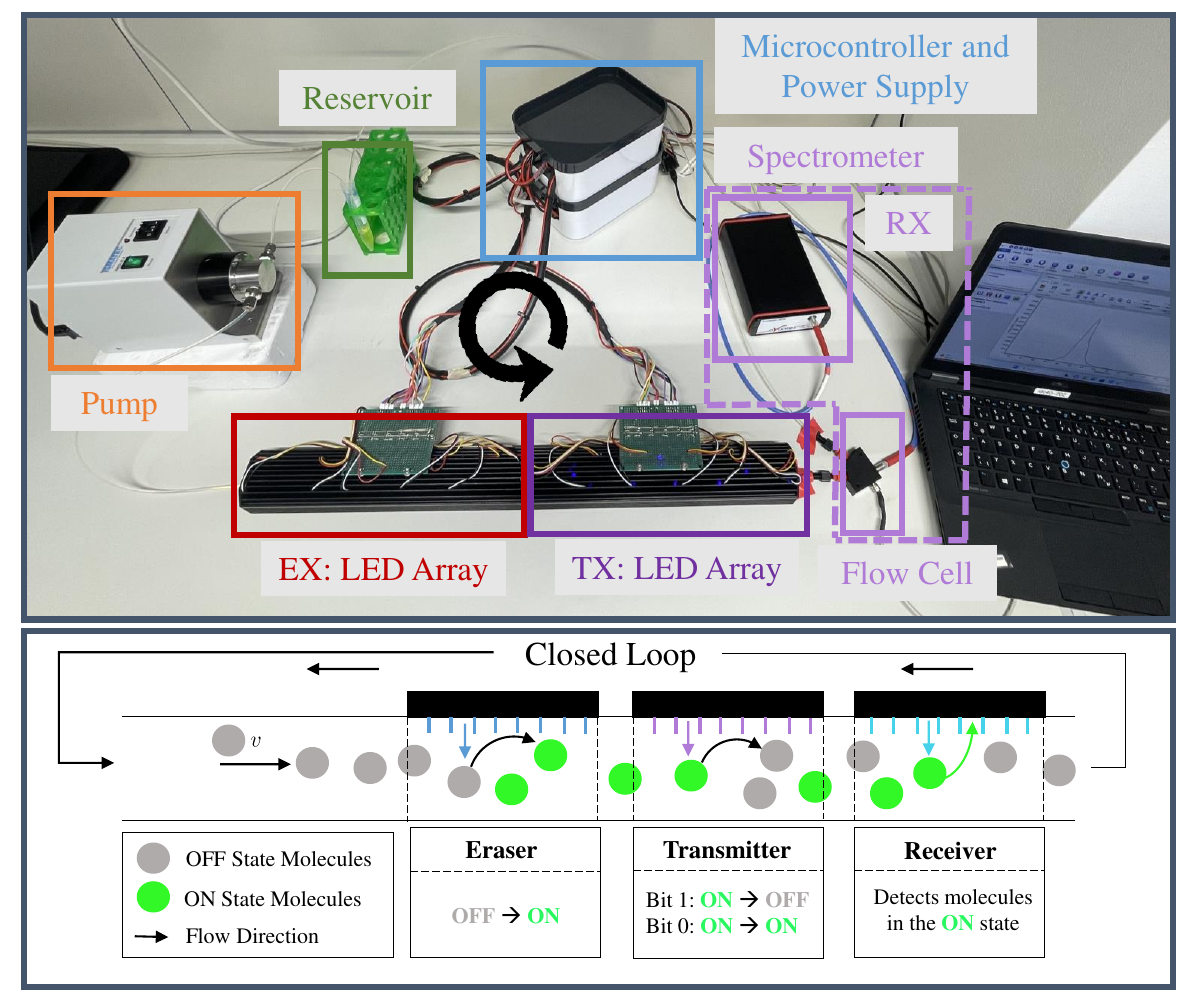}\vspace*{-0.4cm}
  \caption{\textit{Top}: Overview of the building blocks of the experimental closed loop \ac{MC} setup using \ac{GFPD} as signaling molecule. The setup comprises a reservoir, a pump, connecting tubes, \ac{LED} arrays for \ac{EX} and TX, and the \ac{RX} which consists of an \ac{LED} connected to a fluorescence flow cell and a spectrometer. The \ac{LED} arrays are controlled via a microcontroller and are supplied with power by cable. \textit{Bottom}: Schematic representation of the tube containing \ac{GFPD} dissolved in fluid. The \acp{GFPD} can be switched to the ON state and to the OFF state by the \ac{EX} and \ac{TX}, respectively, while the state of \ac{GFPD} can be determined at the \ac{RX}.}
  \label{fig:GFPD Testbed}
  \vspace*{-0.6cm}
\end{figure}

In this section, we introduce the testbed. \Figure{fig:GFPD Testbed} shows the entire testbed (top panel) as well as a schematic representation of its working principle (bottom panel). We highlight the different building blocks of the testbed by colored boxes in \Figure{fig:GFPD Testbed}.
\version{If not noted otherwise, the default parameter values used in our experiments are given in \Table{Table:Parameter}.
\begin{table}[!tbp]
\centering
\caption{Default Values for Experiment.}
\label{Table:Parameter}
\vspace*{-0.25cm}
\resizebox{\columnwidth}{!}{%
\centering
\begin{threeparttable}
{\def\arraystretch{1.3}\tabcolsep=0.2cm
 \begin{tabular}{|l | c | c | r|}
   \hline
 Parameter & Description & Value & Ref. \\ [0.1cm]
 \hline\hline
 \multicolumn{4}{|c|}{Testbed Setup}\\
 \hline\hline
 $r_\mathrm{T}$ & Tube radius & $8 \, \times 10^{-4} \,\si{\meter}$ & \cite{hall2020guyton} \\
 \hline
 $L_\mathrm{T}$ & Tube length & $2.7 \,\si{\meter}$ &  \\
 \hline
 $\volTube$ & Tube volume & $ \pi r_{\mathrm{T}}^2 L_\mathrm{T} = 5.43 \,\si{\milli\liter}$ &  \\
 \hline
 $V_\mathrm{R}$ & Reservoir volume & $2.50 \,\si{\milli\liter}$ & \\
 \hline
 $V_\mathrm{PF}$ & Volume of pump and flow cell & $1.07 \,\si{\milli\liter}$ &  \\
 \hline
  $V_\mathrm{S}$ & Total volume of system & $V_\mathrm{S} = V_\mathrm{PF} + V_\mathrm{R} + \volTube = 9.00 \,\si{\milli\liter}$ &  \\
 \hline
 $Q$ & Volume flux & $9.45 \,\si{\milli\liter \per\minute}$ &  \\
  \hline
 $v_{\mathrm{eff}}$ & Effective flow velocity & $0.078 \,\si{\meter \per\second}$ &  \\
 \hline
 $T_\mathrm{L}$ & Average loop duration & $41 \,\si{\second}$ & \\
  \hline
 $L_\mathrm{EX}$ \&  $L_\mathrm{TX}$ & \ac{EX} \& \ac{TX} length & $0.29 \,\si{\meter}$ & \\
 \hline
  $L_\mathrm{RX}$ & \ac{RX} length & $3 \, \times 10^{-3} \,\si{\meter}$ & \\
  \hline
   $d_{\mathrm{TX},\mathrm{RX}}$ & Distances between end of \ac{TX} and start of \ac{RX} & $ 0.06 \,\si{\meter}$, $ 0.63 \,\si{\meter}$& \\
  \hline
  $\Ti$ & Irradiation duration & $10 \,\si{  \second}$& \\
  \hline
  $T_\mathrm{G}$ & Guard interval & $0 \,\si{\second}, 20 \,\si{\second}$ & \\
  \hline
  $\Ts$ & Symbol duration &  $10 \,\si{\second}, 30 \,\si{\second}$  & \\
  \hline
  $\Delta t$ & Sampling time step & $2\,\si{\second}$ & \\
 \hline\hline
 \multicolumn{4}{|c|}{GFPD Specific Parameter}\\
 \hline\hline
 $\con$ & Concentration of GFPD & $0.3 \,\si{\milli\gram \per\milli\liter}$ & \\
 \hline
  $M_{\mathrm{GFPD}}$ & Molar mass & $26900 \,\si{\gram \per\mol}$ &  \cite{brakemann2011reversibly} \\
 \hline
 $\lambda_\mathrm{OFF}$ & Wavelength for switching OFF${}^{(\textnormal{a})}$ / ON${}^{(\textnormal{b})}$ & $405 \,\si{\nm}$ / $365 \,\si{\nm}$ & \cite{brakemann2011reversibly} \\
 \hline
 $\lambda_\mathrm{T}$ & Wavelength for fluorescence excitation${}^{(\textnormal{c})}$ & $500 \,\si{\nm}$ & \cite{brakemann2011reversibly} \\
 \hline
 $\lambda_\mathrm{E}$ & Fluorescence emitting wavelength & $529 \,\si{\nm}$ & \cite{brakemann2011reversibly} \\
 \hline
 $T_{1/2}$ & Half time in OFF state & $600 \,\si{\second}$ & \cite{brakemann2011reversibly} \\
 \hline
 \end{tabular}
 }
 \begin{tablenotes}[flushleft]\footnotesize\setlength\itemsep{0cm}
\item[(a)] Used at \ac{TX}. 
\item[(b)] Used at \ac{EX}. 
\item[(c)] Used at \ac{RX}. 
\end{tablenotes}
 \vspace{-0.5cm}
\end{threeparttable}
}
\end{table}
}{Due to space constraints, in this paper, we cannot show a table summarizing all default parameter values used. The reader is referred to the corresponding table in the \textit{Arxiv} version of this paper \cite{brand2023closed}.}

\scaleSubsection
\subsection{Experiment Preparation}\label{sec:Testbed_setup}
\scaleSubsectionBelow

\scaleSubsubsection
\subsubsection{GFPD Solution}
\scaleSubsubsectionBelow
In all experiments, we use $V_\mathrm{S} = 9 \, \si{\milli\liter}$ of \ac{GFPD} solution as medium. The \ac{GFPD} solution is composed of a buffer solution with \acp{GFPD} uniformly dissolved in the buffer at a concentration of $\con = 0.3 \,\si{\milli\gram \per\milli\liter}$. 

\scaleSubsubsection
\subsubsection{Tube System, Reservoir, and Pump}
\scaleSubsubsectionBelow
As the first step in all experiments, we fill the tube, the reservoir, and the segments within the pump and the flow cell with $\volTube = 5.43 \,\si{\milli\liter}$, $V_\mathrm{R} = 2.50 \,\si{\milli\liter}$, and $V_\mathrm{PF} = 1.07 \,\si{\milli\liter}$ of the \ac{GFPD} solution, respectively.
For the tube, we use a fluorinated ethylene propylene (FEP) tube with radius $r_\mathrm{T} = 8 \, \times 10^{-4} \, \si{\meter}$ and total length $L_\mathrm{T} = 2.7 \,\si{\meter}$.
The reservoir is realized by a Falcon\textsuperscript{\tiny\textregistered} tube and needed in our system to close the circuit, i.e., both ends of the tube are immersed in the fluid within the reservoir and are hereby connected.
During operation, the \ac{GFPD} solution is pumped through the tube and the reservoir using a gear pump generating an effective volume flux of $Q = 9.45 \,\si{\milli\liter \per\minute}$, which results in an effective flow velocity of $v_{\mathrm{eff}} = 0.078 \,\si{\meter \per\second}$ for the given tube radius. Hence, the average duration $T_\mathrm{L}$, that one \ac{GFPD} molecule needs to circulate one time through the testbed, can be approximated as $T_\mathrm{L} = \frac{V_\mathrm{PF} + \volTube}{Q} = 41 \,\si{\second}$, assuming, on average, no additional delay induced by the~reservoir.

\scaleSubsection
\subsection{Information Transmission}\label{sec:transmission}
\scaleSubsectionBelow
While the pump transports the \acp{GFPD} within the circuit as described above, information can be modulated from outside the tube into the state of the \acp{GFPD} by an \ac{LED}-based \ac{TX}. Later, the \ac{RX} reads out the state of the \acp{GFPD} via fluorescence triggered by an \ac{LED}. Additionally, an \ac{LED}-based \ac{EX} can be used to erase the modulated information. This modulation method is referred to as media modulation and schematically represented in the bottom part of \Figure{fig:GFPD Testbed}.

The \acp{LED} of the \ac{TX}, \ac{EX}, and \ac{RX} are controlled by an ESP32-S2 microcontroller.
In the following, the practical realization and use of \ac{TX}, \ac{RX}, and \ac{EX} are presented in detail. Additionally, for easier understanding, we illustrate the discussion of the functions of the three components with measurement results.

\begin{figure}[!tbp]
    \begin{minipage}[t]{0.48\textwidth}
      \centering
      \includegraphics[width = 0.93\textwidth]{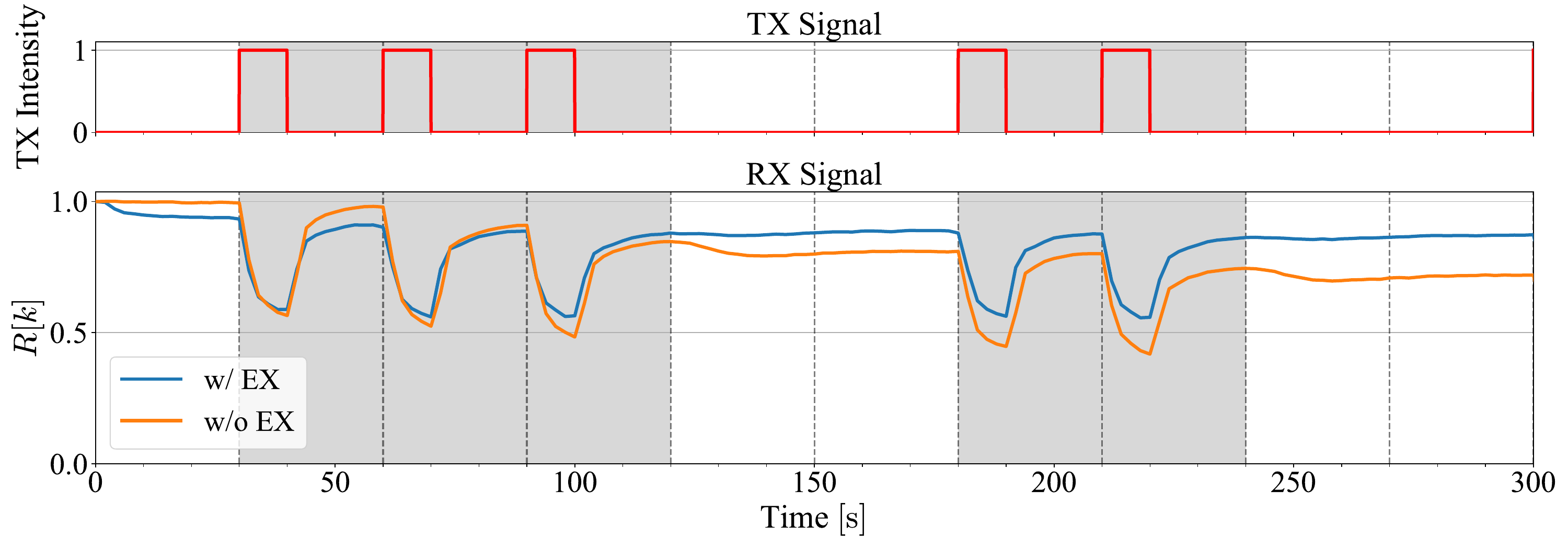}\vspace*{-0.4cm}
      \caption{\textit{Setting}: $\Ti = \SI{10}{\second}, \Tg = \SI{20}{\second}$. \textit{Top}: Normalised \ac{TX} intensity (red) for the transmission of a bit sequence. \textit{Bottom}: The received fluorescence intensity of the system at $\lambda_\mathrm{E} = 529 \, \si{\nm}$ with (blue) and without (orange) the use of an \ac{EX}, respectively. The gray shaded symbol intervals highlight bit 1 transmissions.}
      \label{fig:transmission_20s}
      \vspace*{-0.6cm}
    \end{minipage}
\end{figure}

\scaleSubsubsection
\subsubsection{Transmitter}\label{Transmitter}
\scaleSubsubsectionBelow
The \ac{TX} is used to modulate binary information via media modulation into the state of the \acp{GFPD}.
The \ac{TX} comprises an \ac{LED} array with 12 equally spaced \acp{LED} which can emit light at a wavelength of $\lambda_\mathrm{OFF} = 405 \, \si{\nano\meter}$. The \ac{LED} array is attached to an aluminium circuit board. The board has a total length of $L_\mathrm{TX} = 29 \,\si{\cm}$ and is fixed to a passive cooling element to increase the \ac{LED} operation life time. Moreover, the \acp{LED} are individually wired to guarantee maximum flexibility of our testbed. In the long run, this allows the individual control of each \ac{LED} during the experiments. However, in this work, all \acp{LED} are used simultaneously.

To transmit information, for bit 1, the \ac{TX} irradiates with maximum intensity for an irradiation duration of $T_\mathrm{I}$ to trigger the switching to the OFF state of the available \ac{GFPD} molecules, i.e., molecules which are below the \ac{TX} module during irradiation. For a bit 0, the \ac{TX} is off during $T_\mathrm{I}$. $T_\mathrm{I}$ is followed by a guard interval of duration $T_\mathrm{G}$, during which the \ac{TX} is always inactive. This results in a symbol duration of $\Ts = T_\mathrm{I} + T_\mathrm{G}$.

The top panel of \Figure{fig:transmission_20s} shows the normalised \ac{TX} intensity (red line) as a function of time, which is used to transmit the exemplary bit sequence $\langle 0, 1, 1, 1, 0, 0, 1, 1, 0, 0\rangle$ with $\Ti = \SI{10}{\second}, \Tg = \SI{20}{\second}$, i.e., $\Ts = \SI{30}{\second}$. In \Figure{fig:transmission_20s}, bit 1 transmissions are shaded in gray. For all our experiments in this paper, we use $\Ti = 10 \,\si{\second}$. As guard intervals, $T_\mathrm{G} = 0 \,\si{\second}$ and $ T_\mathrm{G} = 20 \,\si{\second}$ are considered, which results in $\Ts = \SI{10}{\second}$ and $\Ts = \SI{30}{\second}$, respectively.

\scaleSubsubsection
\subsubsection{Receiver}\label{Receiver}
\scaleSubsubsectionBelow
The \ac{RX} is used to read out the current states of the \acp{GFPD}.
To this end, the \ac{RX} is placed at a distance of $d_{\mathrm{TX},\mathrm{RX}}$ after the end of the \ac{TX}. Unless noted otherwise, we use $d_{\mathrm{TX},\mathrm{RX}} =  0.06 \,\si{\meter}$. The \ac{RX} is comprised of a fluorescence flow cell (FIA-SMA-FL-ULT - Ocean Optics) equipped with a $\lambda_\mathrm{T} = 500 \, \si{\nm}$ \ac{LED} used to trigger the fluorescence of \ac{GFPD}. The emitted light upon fluorescence is guided by an optical fiber to a spectrometer (Avantes), where the fluorescence intensity is recorded using AvaSoft 8. 
We denote the received signal after isolating the part of the signal close to the wavelength of interest, i.e., $\lambda_\mathrm{E} = 529 \, \si{\nano\meter}$, as $R[k]$. The discrete time signal $R[k]$ collects the fluorescence intensities measured at the \ac{RX} at time instances $t_k = k \Delta t$ with $k = {0, 1, 2, ..., k_{\mathrm{max}}}$, where $\Delta t$ denotes the time difference between two samples. $k_{\mathrm{max}} = \frac{T_{\mathrm{rec}}}{\Delta t}$ depends on the absolute duration $T_{\mathrm{rec}}$ of the signal recorded by the spectrometer. In this work, we use $\Delta t = 2\,\si{\second}$ unless noted otherwise.

The bottom panel of \Figure{fig:transmission_20s} shows $R[k]$ (orange line) as a function of time in response to the \ac{TX} irradiation intensity. Note that, in this case, no \ac{EX} is used, which is indicated by the label "w/o EX". $R[k]$, as expected, shows clear fluorescence intensity drops for bit 1 transmissions. Conversely, during a bit 0 transmission, the $R[k]$ shows no drop and remains at a high level. Note that we can already see here that the exact shape of $R[k]$ for the current bit depends on the previously sent bits, indicating the existence of \ac{ISI}, which is discussed in detail in \Section{subsec:isi}.

\scaleSubsubsection
\subsubsection{Eraser}\label{Eraser}
\scaleSubsubsectionBelow
The \ac{EX} can be used to reset the state of the \acp{GFPD} to the ON state.
The \ac{EX} component comprises an \ac{LED} array with 12 equally spaced \acp{LED} operating at $\lambda_\mathrm{ON} = 365 \, \si{\nano\meter}$ and has a total length of $L_\mathrm{EX} = 29 \,\si{\cm}$. It is placed upstream next to the \ac{TX}.
When we use the \ac{EX}, unless otherwise noted, the \ac{EX} is turned on at $t=0\, \si{\second}$, i.e., for these experiments the \ac{EX} is continuously active. Hereby, it enables resetting the \acp{GFPD} from the OFF state back to the ON state. Hence, with the \ac{EX} turned on, the number of available ON state molecules at the \ac{TX} is increased.

\Figure{fig:transmission_20s} also shows $R[k]$ as a function of time for an active \ac{EX} (blue line), labeled as "w/ EX". We observe that still clear peaks are visible in $R[k]$ for bit 1 transmissions. However, with the \ac{EX}, the received signal for bit 0 transmissions is almost constant, which is desired\footnote{In fact, for the first bit 0, \Figure{fig:transmission_20s} shows a slight drop of $R[k]$ which is not observed for experiments without the \ac{EX}. We explain the drop with the \ac{EX} with the switching of the \ac{GFPD} molecules from the equilibrium state, cf. Footnote 1 in \Section{sec:GFPD_photoswitchable}, to the slightly less fluorescent ON state at the beginning of the experiment.}.

%% file: characterization.tex
In this section, we mathematically model the received signal and characterize the different forms of \ac{ISI} present in the testbed.
\vspace{-0.4cm}
\scaleSubsection
\subsection{Mathematical Description of the Received Signal and the Pilot Sequence}
\label{received_signal_math}
\scaleSubsectionBelow
In the investigated experimental \ac{MC} system, $N_{\mathrm{Sym}}$ binary symbols $b[i] \in \{0, 1\}$, $i = {0, 1, 2, ..., N_{\mathrm{Sym}}-1}$, are transmitted and signal $R[k]$ is received. $R[k]$ can be partitioned as follows
\scaleAlign
\begin{align}
 \mathcal{R}_i[k] = R\left[k + i \frac{\Ts}{\Delta t}\right] \qquad \mathrm{for} \quad 0 \leq k< \frac{\Ts}{\Delta t} \;, \label{eq:subsignals}
\end{align}
where signal $\mathcal{R}_i[k]$ corresponds to symbol $b[i]$. For detection, we use samples $\rec{i} = \mathcal{R}_i[k_{\mathrm{p}}]$, i.e., we use one sample $\rec{i}$ per symbol interval $i$, which corresponds to the sampling instant $k_{\mathrm{p}}$, where the local minimum for a bit 1 transmission is expected. The value of $k_{\mathrm{p}}$ is experimentally determined based on a pilot sequence $b^{\mathrm{P}}$ of length $\nPilots$ given by $b[i]$, $i \in \{0, 1, ..., \nPilots-1\}$, which is sent at the beginning of each transmission and is known at the \ac{RX}. In particular, to obtain $k_{\mathrm{p}}$, we take the median of the indices of the local minima of all bit 1 transmissions in $b^{\mathrm{P}}$. For the experiments shown in this work, we use the pilot sequence $b^{\mathrm{P}} = \langle 0, 1, 1, 1, 0, 0, 1, 1, 0, 0\rangle$, of length $\nPilots = 10$, which is identical to the bit sequence used in \Figure{fig:transmission_20s}.

\scaleSubsection
\subsection{Channel Symbol Responses}
\label{subsec:sir}
\begin{figure}[t!]
    \centering
    \begin{minipage}[t]{0.48\textwidth}
      \centering
      \includegraphics[width = 0.92\textwidth]{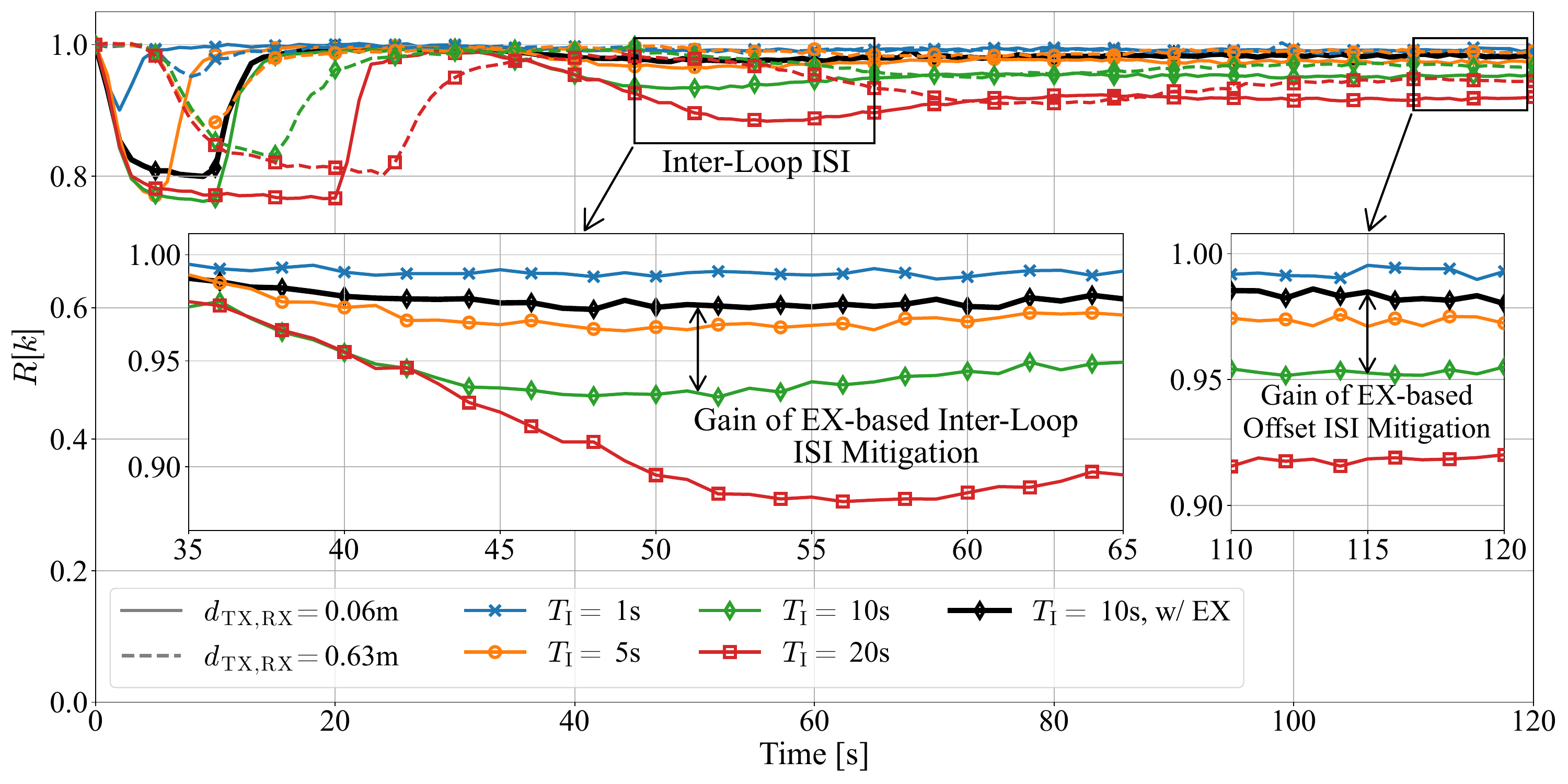}\vspace*{-0.4cm}
      \caption{Recorded fluorescence intensity signal $R[k]$ as a function of time for different irradiation durations $\Ti$ and \ac{TX}-\ac{RX} distances $d_{\mathrm{TX},\mathrm{RX}} = 0.06 \,\si{\meter}$ and $d_{\mathrm{TX},\mathrm{RX}} = 0.63 \,\si{\meter}$. Here, we use $\Delta t = 1\,\si{\second}$.}
    \label{fig:experimental-cir}
    \end{minipage}
    \hfill
  \vspace{-0.6cm}
\end{figure}

\scaleSubsectionBelow
\Figure{fig:experimental-cir} shows the received signals for different \ac{TX} and channel settings for a single bit 1 transmission. In particular, we vary the \ac{TX} irradiation duration $\Ti$ and $d_{\mathrm{TX},\mathrm{RX}}$. Increasing $\Ti$ also increases the fluorescence intensity drop until for $\Ti = 5 \, \si{\second}$ and  $\Ti = 10 \, \si{\second}$ $R[k]$ exhibits a minimum for $d_{\mathrm{TX},\mathrm{RX}} = 0.06 \,\si{\meter}$ and $d_{\mathrm{TX},\mathrm{RX}} = 0.63 \,\si{\meter}$, respectively. Increasing $\Ti$ further does not change the minimum fluorescence intensity as the average energy that a \ac{GFPD} molecule receives is limited. In particular, the average energy received by a \ac{GFPD} is dependent on the irradiation power $P$ of the \acp{LED} at the \ac{TX}, the average flow velocity $v_{\mathrm{eff}}$, the length of the \ac{TX} $L_{\mathrm{TX}}$, and $\Ti$. A detailed investigation of this aspect is left for future work.

Moreover, we see in \Figure{fig:experimental-cir} that increasing $d_{\mathrm{TX},\mathrm{RX}}$ from $d_{\mathrm{TX},\mathrm{RX}} = 0.06 \,\si{\meter}$ to $d_{\mathrm{TX},\mathrm{RX}} = 0.63 \,\si{\meter}$ results in a less severe fluorescence intensity drop, as the modulated \acp{GFPD} have more time to spontaneously switch back to the ON state and disperse further during the propagation from \ac{TX} to \ac{RX}.

Finally, \Figure{fig:experimental-cir} shows also one result for the case with \ac{EX}. Here, the \ac{EX} is started a few minutes before $t=0 \,\si{\second}$. We observe that, with the \ac{EX}, the drop in $R[k]$ is smaller compared to the case without the \ac{EX}. This can be explained by the slightly smaller difference between ON state and OFF state compared to the difference between equilibrium state and OFF state, cf. Footnote 1.
The different forms of \ac{ISI} are discussed in the next section.

\scaleSubsection
\subsection{ISI Characterization and Mitigation}
\label{subsec:isi}
\scaleSubsectionBelow
We identified four different forms of \ac{ISI} in our testbed, with three of them appearing only for long term experiments in closed loop systems. These forms of \ac{ISI} occur on different time scales and are discussed in the following.

\scaleSubsubsection
\subsubsection{Channel ISI}
\label{subsubsec:channel_isi}
\scaleSubsubsectionBelow
During their propagation within the channel from \ac{TX} to \ac{RX}, signaling molecules from consecutive symbols may overlap, leading to \ac{ISI}. To differentiate between different forms of \ac{ISI}, we refer to this form as \textit{channel \ac{ISI}}. Channel \ac{ISI} has been observed in existing \ac{MC} testbeds, e.g., \cite{grebenstein2018biological, wang2020understanding}. We refer to \cite{wang2020understanding} for a comprehensive discussion of channel~\ac{ISI}.

Channel \ac{ISI} can be mitigated by using a guard interval, during which the \ac{TX} is inactive and the signaling molecules have time to be transported out of the channel before the next symbol is transmitted. Of course, the success of the \ac{ISI} mitigation depends on the guard interval duration $\Tg$, and improves for increasing $\Tg$. However, increasing $\Tg$ decreases the data rate, i.e., there exists a tradeoff between the performance of channel \ac{ISI} mitigation and the data rate. Note that also without a guard interval, the impact of channel \ac{ISI} (and not channel \ac{ISI} itself) on the sample $d[i]$ may vary dependent on the symbol duration $\Ts$ and $k_{\mathrm{p}}$.

\scaleSubsubsection
\subsubsection{Inter-Loop ISI}
\label{subsubsec:loop_isi}
\scaleSubsubsectionBelow
Due to the closed loop, a second form of \ac{ISI} occurs, which we call \textit{inter-loop \ac{ISI}}. As the signaling molecules remain in the system, they may interfere not only with neighbouring symbols but also with symbols transmitted much later. In particular, \ac{GFPD} molecules, which are switched to the OFF state during a bit 1 transmission, may travel several times through the loop before they spontaneously switch back to the ON state. Hereby, these signaling molecules likely impact the received signal multiple times, while the distinctness and the number of recurrences depend on the length of the loop and the half-life of the OFF state $T_{1/2}$. The left inset of \Figure{fig:experimental-cir} shows that inter-loop \ac{ISI} is visible as a dip in the received signal and can be reduced by utilizing an \ac{EX}. Note that the first inter-loop \ac{ISI} peak appears approximately $35 \,\si{\second}$ after the main peak, which is close to the calculated loop duration of $T_\mathrm{L}$. Furthermore, \Figure{fig:experimental-cir} shows that inter-loop \ac{ISI} becomes increasingly severe as $\Ti$ increases.

\scaleSubsubsection
\subsubsection{Offset ISI}
\label{subsubsec:offset_isi}
\scaleSubsubsectionBelow
Since the OFF state \acp{GFPD} disperse within the system over the loops, the \ac{ISI} gradually transforms from inter-loop \ac{ISI}, characterized by distinctive dips, to a temporal offset of the fluorescence intensity, which we denote as \textit{offset \ac{ISI}}. The offset \ac{ISI} is visible in \Figure{fig:experimental-cir} as an overall lower fluorescence intensity level compared to the initial level. In particular, for strong offset \ac{ISI}, the received fluorescence intensity in response to a bit 0 transmission is reduced, and therefore closer to the intensity in response to a bit 1 transmission. This affects the reception performance. The intensity of offset \ac{ISI} depends on the number of bit 1 transmissions, symbol duration $\Ti$, and half-life $T_{1/2}$, and can be mitigated by using an \ac{EX}. The right inset of \Figure{fig:experimental-cir} shows that offset \ac{ISI}, visible as the difference between $R[k]$ and the initial fluorescence intensity, increases for increasing $\Ti$ and decreases when utilizing the \ac{EX}.

\scaleSubsubsection
\subsubsection{Photobleaching and Permanent ISI}
\label{subsubsec:permanent_isi}
\scaleSubsubsectionBelow
Finally, after intensive irradiation, \ac{GFPD} molecules may be permanently degraded. This effect occurs in long experiments and is referred to as photobleaching. While the decrease in fluorescence intensity due to offset \ac{ISI} is reversible, photobleaching results in a long term and permanent decrease of \ac{GFPD} fluorescence.
As \acp{GFPD} are irradiated by the \ac{EX}, the \ac{RX}, and the \ac{TX} in the testbed, all three light sources contribute to photobleaching. Therefore, we denote the portions caused by the \ac{TX}, by the \ac{EX}, and by the \ac{RX} as \textit{permanent \ac{ISI}}, \textit{eraser bleaching}, and \textit{measurement bleaching}, respectively, where only the permanent \ac{ISI} is dependent on the transmitted signal\footnote{Note that only the \ac{TX} irradiation varies as a function of the transmitted sequence. Therefore, only the photobleaching caused by the \ac{TX} can be interpreted as \ac{ISI}, whereas the photobleaching caused by the \ac{EX} and the \ac{RX} is independent of the transmitted sequence.}.
Photobleaching affects the availability of \ac{GFPD} for media modulation and hereby impacts the transmission of future symbols. Photobleaching increases with the irradiation energy, e.g., for higher irradiation power and longer irradiation duration \cite[Suppl. Fig. 7]{brakemann2011reversibly}. Therefore, there exists a tradeoff between switching accuracy (which increases with the irradiation energy provided by the \ac{EX} and \ac{TX}) and photobleaching. We observe that photobleaching limits the maximum duration of our experiments. As it manifests itself only after many transmission cycles, it is not visible in~\Figure{fig:experimental-cir}.

\scaleSubsubsection
\subsubsection{ISI Mitigation}
\label{subsubsec:isi_mitigation}
\scaleSubsubsectionBelow
To overcome inter-loop and offset \ac{ISI}, we exploit the spontaneous switching property of \ac{GFPD} and the \ac{EX} for passive and active inter-loop and offset \ac{ISI} mitigation, respectively. 
In fact, \acp{GFPD} spontaneously switch back from the OFF state to the ON state with a half-life of $T_{1/2} = 600 \,\si{\second}$. Hereby, information modulated to the state of the \acp{GFPD} is erased naturally, and offset \ac{ISI} is reduced. 
In addition, the \ac{EX} actively triggers the switching to the ON state. If the \ac{EX} is ideal, it switches all traversing \acp{GFPD} from the less fluorescent OFF state to the fluorescent ON state. Of course, in reality, only a fraction of the \acp{GFPD} may be switched because the power of the \acp{LED} is limited. \Figure{fig:experimental-cir} reveals the success of \ac{EX}-based \ac{ISI} mitigation for $\Ti = 10 \,\si{  \second}$. In particular, we see that both inter-loop \ac{ISI} and offset \ac{ISI} are reduced, which increases the overall fluorescence intensity by approximately 4\%\footnote{Note that the forms of \ac{ISI} mitigation described here impact the signaling molecules themselves. Hence, unlike equalization schemes, they are not part of the detection scheme at the \ac{RX} and therefore do not contribute to the computational complexity needed for detection.}.
\vspace{0.05cm}

%% file: performance.tex
In the following, we describe two different threshold detection schemes. In all cases, the pilot sequence $b^{\mathrm{P}}$ is used to determine the optimal threshold value. However, the detection techniques differ in terms of complexity. The \ac{BD}, presented in \Section{basic_threshold_detection}, uses symbol-by-symbol detection, i.e., each symbol is detected independent of the other symbols. In contrast, the \ac{DD} introduced in \Section{differential_detection} exploits two consecutive samples.

\scaleSubsection
\subsection{Basic Threshold Detector}\label{basic_threshold_detection}
\scaleSubsectionBelow
For the \ac{BD}, the threshold $\threshR$ is used to estimate the transmitted symbol as follows
\scaleAlign
\begin{equation}
    \bitest[\mathrm{R}]{i} = \begin{cases} 0, \qquad &\text{if } \rec{i} \geq \threshR \\
         1, \qquad &\text{otherwise}
    \end{cases}\;.
\label{eq:td}
\end{equation}

To find a suitable threshold, we choose the threshold value $\threshR$ that minimizes the Hamming distance between the symbols of pilot sequence $b^{\mathrm{P}}$ and their estimates. If the choice of $\threshR$ is ambiguous, i.e., more than one $\threshR$ minimizes the Hamming distance, we select the lowest of the possible values.
Once $\threshR$ is computed from the pilot sequence, the symbols $\bitest[\mathrm{R}]{i}$ for $i \in \{\nPilots, ..., \nSymb-1\}$ are computed based on \Equation{eq:td}.

\scaleSubsection
\subsection{Differential Threshold Detector}\label{differential_detection}
\scaleSubsectionBelow
Another threshold based approach is the \ac{DD} scheme. \ac{DD} was proposed for \ac{MC} in \cite{li2015low} and implemented for an \ac{MC} testbed in \cite{grebenstein2018biological}. In fact, in \cite{grebenstein2018biological}, the authors have shown the robustness of \ac{DD} against a slow drift in the received signal, which motivates the usage of such a detector also in this work.

For \ac{DD}, the difference $d'[i]$ between the current sample and the sample of the previous symbol is utilized. $d'[i]$ is defined as follows
\scaleAlign
\begin{equation}
    \recDiff{i} = \begin{cases}
        0, \qquad &\text{for } \, i = 0 \\
        \rec{i} - \rec{i-1}, \qquad &\text{otherwise}
    \end{cases}\;.
\end{equation}

If the preceding and the current symbol are equal, the difference between the sampling values, i.e., $\recDiff{i}$ is likely to be small in its absolute value. If a change in transmitted symbols occurs, then $\recDiff{i}$ will more likely have a large absolute value. In this case, if a bit 0 is transmitted after a bit 1, we expect $\recDiff{i} > 0$ and if a bit 1 follows a bit 0, $\recDiff{i} < 0$. These considerations are exploited in the decision rule of \ac{DD} as follows
\scaleAlign
\begin{equation}
    \bitest[\mathrm{D}]{i} = \begin{cases}
                    \bitest[\mathrm{D}]{i - 1}, \qquad &\text{if } |\recDiff{i}| \leq \threshD \\
                    \frac{1 - \frac{\recDiff{i}}{|\recDiff{i}|}}{2}, \qquad &\text{otherwise}
                \end{cases} \;.
\end{equation}

To find a suitable value for $\threshD$, we first determine the set of indices of the transmitted pilot symbols that are repetitions of their respective preceding symbols as follows
\scaleAlign
\begin{equation}
    \idxSym{=} = \{i \, | \, \bit{i} = \bit{i - 1}, \; i \in \{0, 1, ..., \nPilots-1\} \}.
\end{equation}
Then, we use this set to find the threshold $\threshD$ as follows
\scaleAlign
\begin{align}
   \threshD = \max_{i \in \idxSym{=}} |\recDiff{i}| \;,
   \label{eq:thresholdDifferential}
\end{align}
i.e., based on our observation that $|d'[k]|$ decreases with increasing experiment duration, we use the maximum here.

%% file: evaluation.tex
In this section, we discuss the end-to-end communication performance of the developed testbed. In particular, we evaluate the \ac{SER} for different transmission lengths $\nSymb$ and different data rates $R$ using the detection schemes discussed in \Section{Estimation_and_Detection}. We show results with and without \ac{EX}-based \ac{ISI} mitigation.
\scaleSubsection
\subsection{Visualizing the Detection Schemes}\label{detection_schemes_real_data}
\scaleSubsectionBelow
\begin{figure}[!tbp]
    \begin{minipage}[t]{0.48\textwidth}
    \centering
      \includegraphics[width = \textwidth]{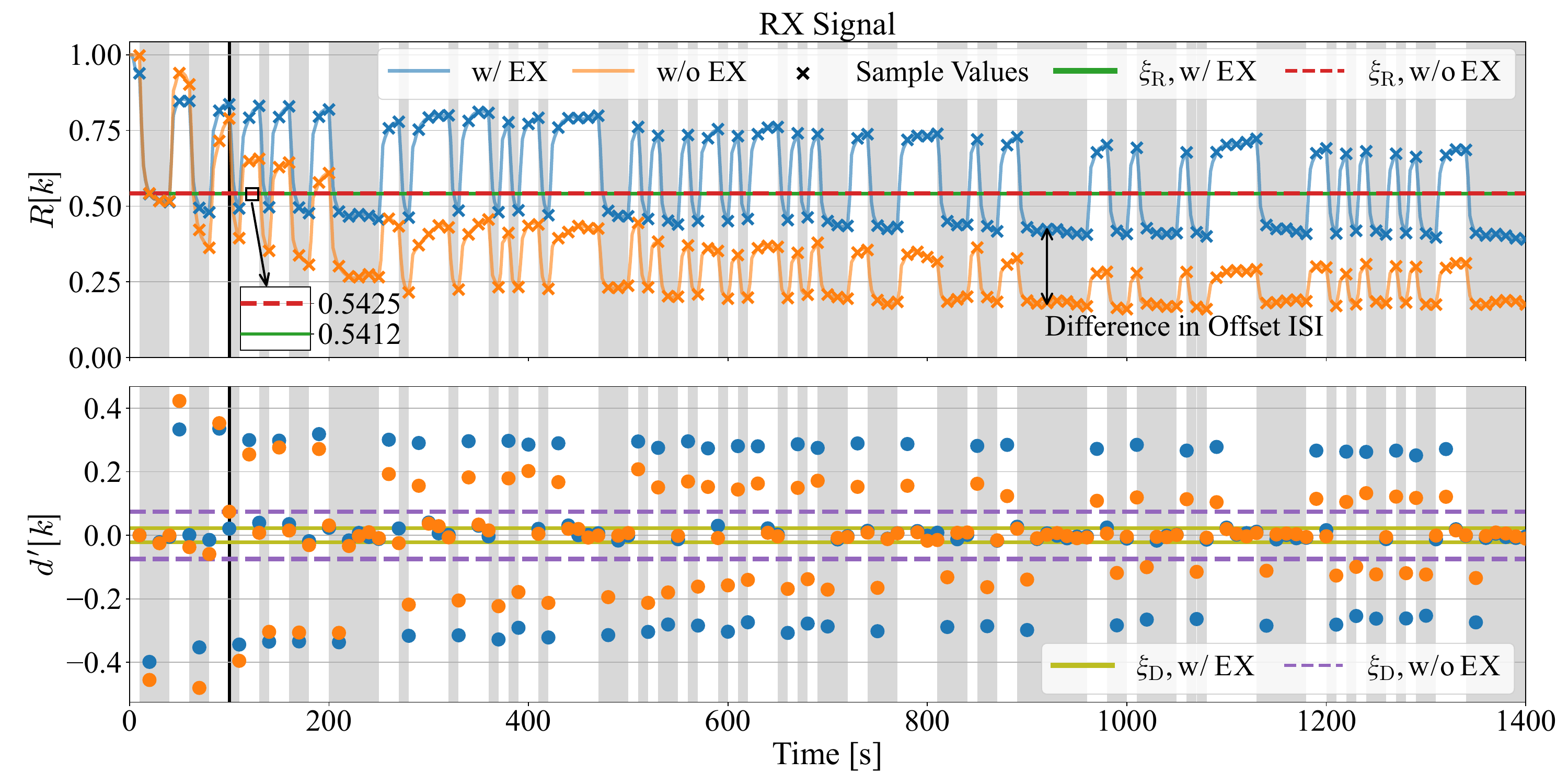}\vspace*{-0.4cm}
      \caption{\textit{Setting}: $\Ti = \SI{10}{\second}, \Tg = \SI{0}{\second}$, from which $R = 6.0 \, \si{\bit \per \minute}$ follows. \textit{Top}: Recorded signal and \ac{BD}. The first segment of the received fluorescence intensity of a $84 \,\si{\minute}$ long experiment and the basic threshold values $\threshR$ as defined in \Section{basic_threshold_detection} as a function of time, respectively. The markers indicate the sampled values used for detection. \textit{Bottom}: \ac{DD} scheme. The differentiated samples (dots) and the differential threshold values $\threshD$ as defined in \Equation{eq:thresholdDifferential} as a function of time. \textit{Both}: The gray-shaded symbol intervals highlight bit 1 transmissions. The vertical line (solid black) indicates the end of the pilot sequence which is employed to determine the threshold values.}
      \label{fig:detection}
      \vspace*{-0.6cm}
    \end{minipage}
\end{figure}
The top panel of \Figure{fig:detection} shows the normalized fluorescence intensity (solid orange and blue lines) over time and the sampled values $\rec{i}$ (markers) using as setting $\Ti = \SI{10}{\second}$ and $ \Tg = \SI{0}{\second}$. In addition, the bottom of \Figure{fig:detection} shows the corresponding difference of two consecutive samples $d'[i]$ (dots).
We further show the threshold values $\threshR$ and $\threshD$, which are derived based on the pilot sequence, whose end is highlighted by a black vertical solid line.

The top panel of \Figure{fig:detection} shows that, even without a guard interval, clear peaks are visible in the received signal for bit 1 transmissions. We observe that the amplitudes of the recorded signals $R[k]$ decrease over time, which is expected. In fact, without the \ac{EX}, the overall fluorescence intensity level decreases fast, which is due to the increasing offset \ac{ISI}, cf. \Section{subsec:isi}. In contrast, when using the \ac{EX}, the decrease is slower, but still exists. We conclude that in this case the decrease is mainly caused by degradation of \ac{GFPD} due to photobleaching, cf. \Section{sec:GFPD}. Moreover, we observe from the bottom panel that the dots, which denote $d'[k]$, approach $d'=0$ over time. This indicates that the difference of the fluorescence intensity levels of bits 0 and bits 1 decreases over time, posing a challenge especially for long term experiments.

\Figure{fig:detection} further shows that $\threshR$ is practically identical for the cases with and without \ac{EX} and is, in this case, determined by the second sample within the pilot sequence of length $\nPilots = 10$. In fact, the second sample appears at a time instant where $R[k]$ has not yet been impacted by the \ac{EX}, which explains the similarity. Moreover, we observe that $\threshD$ used for the case without \ac{EX} is larger compared to that for the case with \ac{EX}. This is intuitive as $R[k]$ for the \ac{EX}-free case varies more, which is reflected in the larger $\threshD$.

\scaleSubsection
\subsection{SER Evaluation}\label{ser_evaluation}
\scaleSubsectionBelow
\begin{figure}[!tbp]
    \begin{minipage}[t]{0.48\textwidth}
    \centering
    \includegraphics[width=0.92\textwidth]{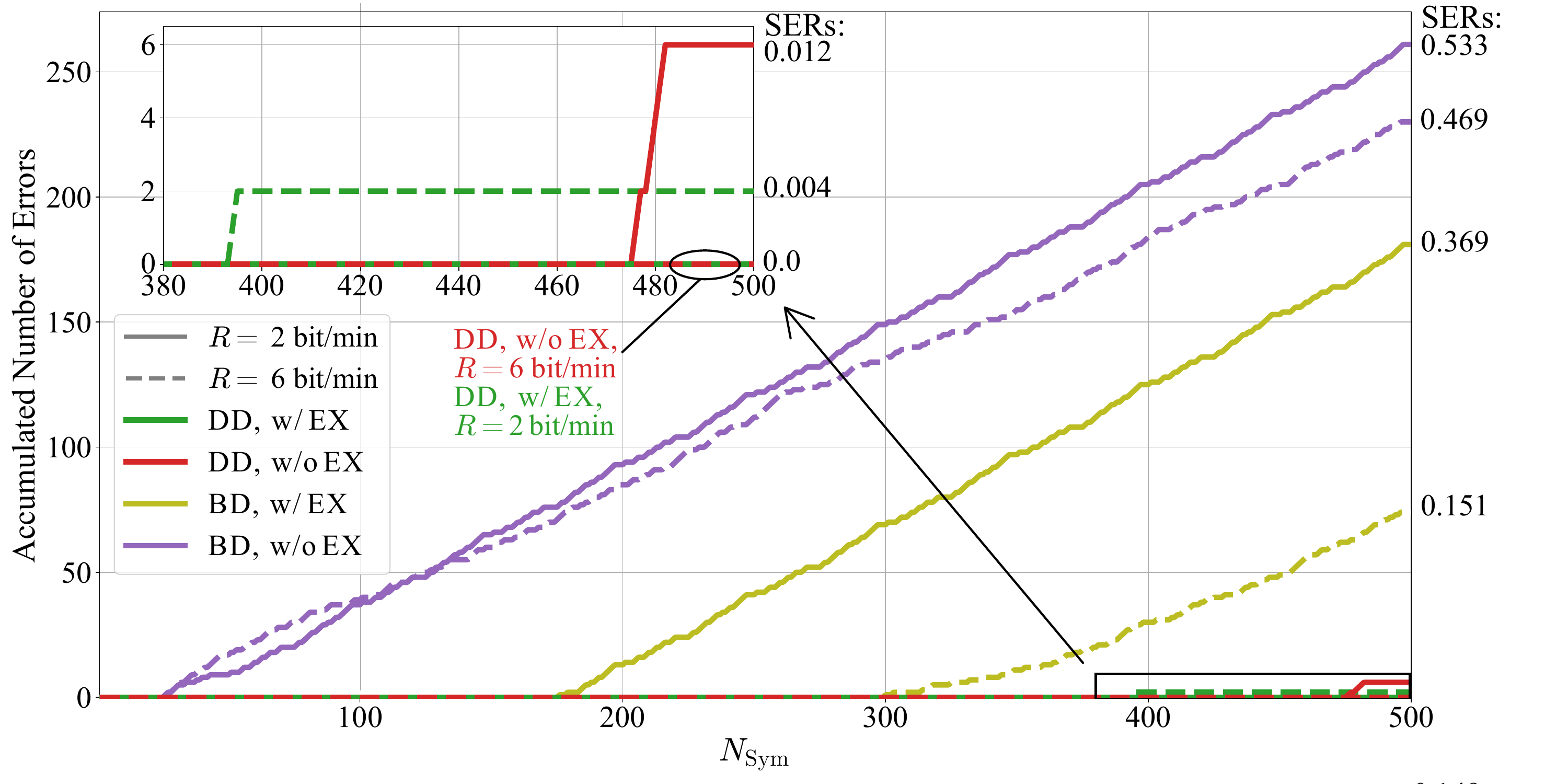}\vspace*{-0.4cm}
    \caption{Detection accuracy: Accumulated number of transmission errors as a function of the transmission sequence length $\nSymb$ for different data rates and detection schemes. Data rates of $R = 2.0 \, \si{\bit \per \minute}$ and $R = 6.0 \, \si{\bit \per \minute}$ are achieved with $\Ti = \SI{10}{\second}, \Tg = \SI{20}{\second}$, cf. \Figure{fig:transmission_20s}, and $\Ti = \SI{10}{\second}, \Tg = \SI{0}{\second}$, cf. \Figure{fig:detection}, respectively. w/ EX and w/o EX indicate the cases with and without \ac{EX} based \ac{ISI} mitigation, respectively.}
    \label{fig:ber_overview}
    \vspace*{-0.6cm}
    \end{minipage}
\end{figure}
\Figure{fig:ber_overview} shows the accumulated number of transmission errors as a function of the transmission sequence length $\nSymb$ for different settings. Hence, in this plot, horizontal lines indicate error free transmission for the length of the line. Additionally, we show the \ac{SER} for $\nSymb = 500$, which is the maximum number of symbols transmitted, on the right-hand-side $y$-axis.

First, we discuss the results for $R = 6.0 \, \si{\bit \per \minute}$. From \Figure{fig:ber_overview}, we observe that the \ac{BD} scheme enables error-free \ac{MC} only at the beginning of the experiment. For long transmission times, the \ac{BD} scheme yields a poorer performance compared to the \ac{DD} scheme. In fact, due to offset \ac{ISI} and photobleaching, the channel state information gained from the pilot sequence and used for the computation of $\threshR$ is eventually outdated, which leads to an increase in errors. This can also be observed in the top panel of \Figure{fig:detection}, where for $t>200 \, \si{\second}$ the orange curve is always below $\threshR$. We observe that, when using the \ac{EX} to mitigate offset \ac{ISI}, the first errors occur later.
We further observe that \ac{DD}, due to its robustness to imperfect channel state information, fails only for two symbols and not at all for the cases with and without \ac{EX}, respectively. In particular, offset \ac{ISI} and photobleaching effects are successfully canceled out when taking the difference between two consecutive samples. We conclude that, when using \ac{DD}\footnote{Note that the slightly higher \ac{RX} complexity of the \ac{DD} compared to the \ac{BD} is negligible.}, there exist system settings, where an \ac{EX} is not essential. 

Finally, we observe that decreasing the data rate from $R = 6.0 \, \si{\bit \per \minute}$ to $R = 2.0 \, \si{\bit \per \minute}$ by using a guard interval of duration $\Tg = \SI{10}{\second}$ increases the error rate, which seems counter intuitive at first. In fact, using $\Tg = \SI{10}{\second}$ is expected to decrease the impact of channel \ac{ISI}, while leaving the other forms of \ac{ISI} unmodified. However, as for $R = 2.0 \, \si{\bit \per \minute}$ the experiment takes three times longer, eraser bleaching and measurement bleaching becomes more severe, which explains the performance loss.

%% file: conclusion.tex
In this work, we introduced an experimental closed loop tube-based testbed, which can be used for \ac{MC} with \ac{GFPD} as signaling molecule.
We showed that media modulation can be utilized to transmit binary data in a closed loop system over a long transmission time duration.
We identified and characterized inter-loop \ac{ISI}, offset \ac{ISI}, and permanent \ac{ISI} as specific forms of \ac{ISI}, which occur for long experiments in closed loop testbeds and appear on different time scales. 
Finally, we proved that our testbed enables error-free signal transmission of $500 \,\si{bit}$ at a data rate of $6.0 \,\si{ \bit \per \minute}$ when using \ac{DD}.

While, in this work, we identified new forms of \ac{ISI} and proposed methods to mitigate them, it would be helpful to understand how to handle possible resulting tradeoffs. For example, analyzing the tradeoff between \ac{EX}-based \ac{ISI} mitigation and eraser bleaching, which limits the operation duration of the testbed, is an interesting topic left for future work. Investigating the possibility of increasing the data rate by using higher-order modulation via different irradiation intensities is another promising direction for further research.